\documentclass[atmp]{ipart_v1}
\usepackage{amsbsy,enumerate}

\Year{2015}
\Vol{0}
\Issue{0}

\firstpage{1}

\usepackage{epsfig}

\numberwithin{equation}{section}

\begin{document}
\title[The entropy excess and entropy excess ratio $\cdots$]{Studying the entropy excess and entropy excess ratio in $^{105,106,107}$Pd within BCS model}

\author[Azam Rahmatinejad, Tayeb Kakavand, Rohallah Razavi]{Azam Rahmatinejad, Tayeb Kakavand, Rohallah Razavi}

\begin{abstract}
Pairing correlations and their influence on nuclear properties has been studied within BCS model. Using this theoretical model with inclusion of pairing interaction between nucleons, nuclear level density and entropy of $^{105,106,107}$Pd have been extracted. The results well coincide with the empirical values of the nuclear level densities obtained by Oslo group. Then the entropy excess of $^{107}$Pd and $^{105}$Pd compared to $^{106}$Pd as a function of the temperature has been studied. Also the role of neutron and proton system in entropy excess have been investigated by the using of the entropy excess ratio proposed by Razavi et al. [R. Razavi, A.N. Behkami, S. Mohammadi, and M. Gholami, Phys. Rev. C 86, 047303 (2012)].
\end{abstract}

\maketitle

\section{Introduction}

Pairing correlations are important in investigation of nuclear properties especially at low temperatures. The BCS model with inclusion of pairing interaction between nucleons inside nuclei is one of the most successful microscopic models for describing nuclear properties with inclusion of paring correlation between nucleons. This theoretical model is based on Bardeen-Cooper- Schrieffer superconducting theory \cite{bcs}. Nuclear level density, which can be interpreted as the number of nuclear levels per MeV of excitation energy, is the starting point to deduce thermodynamic quantities of nuclei. In this work nuclear level densities of $^{105,106,107}$Pd isotopes have been extracted within the BCS model and the results have been compared with the corresponding experimental data that recently obtained by Oslo group \cite{exp.level density}. The present investigation shows good agreement between our results and experimental values.
Extracted entropies within the BCS model for $^{105,106,107}$Pd show an entropy excess in $^{105}$Pd and $^{107}$Pd compared to $^{106}$Pd. This entropy excess has been interpreted as a single hole entropy and single particle entropy, respectively. These entropy excess can be observed in extracted entropies from experimental data on nuclear level densities of the isotopes as it has shown in the Fig.3.
In addition as in the BCS formalism neutron and proton system can be studied separately, one can study neutron and proton system role in entropy excess by the use of entropy excess ratio introduced in our previous work \cite{razavi2013}.
In the present work, the single hole and single particle entropy as a function of temperature have been compared with together. Also, the entropy excess ratio has been studied for neutron and proton system of the $^{107}$Pd and $^{105}$Pd.

\section{Summary of the theory}\label{model}
Using the Hamiltonian describing a system of fermions interacting with pairing force, diagonalized by means of Bogoliubov transformation \cite{moretto}, logarithm of grand canonical partition function can be given by \cite{razavi2012}:

\begin{equation} \label{1}
\Omega(\alpha,\beta)=Tr(-\beta\hat{H}-\lambda\hat{N})=-\beta\sum_k(\epsilon_k-\lambda-E_k)+2\sum_kLn[1+exp(-\beta E_k)]-\beta\frac{\Delta^2}{G},
\end{equation}
where $\epsilon_k$  is the single particle fermion energy and $E_k$  is quasi particle energy. The $G$ and $\Delta$ are the strength of paring and pairing gap parameter, respectively. The chemical potential parameter and statistical temperature have been indicated as $\lambda$ and $T$ $(T=1/\beta)$, respectively.
The most probable value for gap parameter can be determined by the use of the following gap equation \cite{razavi2012}:
\begin{equation} \label{2}
\frac{2}{G}=\sum_k\frac{1}{E_k}\tanh\frac{\beta E_k}{2}.
\end{equation}
\begin{equation} \label{3}
N = \sum_k (1-\frac{\varepsilon_k-\lambda}{E_k}tanh\frac{\beta E_k}{2}),
\end{equation}
where $N$ is the number of nucleons. Eq.2 and Eq.3 can be solved together for the given values of $N$ and $G$ in order to obtain pairing gap and chemical potential as a function of temperature. Using obtained values for $\Delta(T)$ and $\lambda(T)$, excitation energy $(E)$ and Entropy $(S)$ can be extracted:
\begin{equation} \label{4}
E =-\frac{\partial\Omega}{\partial\beta}=\sum_k(1-\frac{\varepsilon_k-\lambda}{E_k}tanh\frac{\beta E_k}{2})\varepsilon_k-\frac{\Delta^2}{G},
\end{equation}
\begin{equation} \label{5}
S=\Omega-\alpha N+\beta E=2\sum_kLn[1+exp(-\beta E_k)]+2\beta\sum_k\frac{E_k}{1+exp(\beta E_k)}.
\end{equation}
The entropy excess , which is interpreted as the single hole and single particle entropy, is defined by \cite{{razavi2013},{razavi2012}}:
\begin{equation}\label{6}
\Delta S(hole)=S(odd A)-S(A+1).
\end{equation}
\begin{equation}\label{7}
\Delta S(particle)=S(odd A)-S(A-1).
\end{equation}

Using neutron and proton entropy excess ratio one can study The role of neutron and proton systems in entropy excess (i=p,n) \cite{{razavi2013},{razavi2012}}:
\begin{equation} \label{8}
R_i=\frac{\Delta S_i}{\Delta S},
\end{equation}
where
\begin{equation} \label{9}
\Delta S_i=S_i(oddA)-S_i(A\pm1).
\end{equation}
Nuclear level density can be deduced by following equation:
\begin{equation} \label{10}
\rho(N,Z,U)=\frac{\omega(N,Z,U)}{(2\pi \sigma^2)^{1/2}},
\end{equation}
where, $\sigma^2$ is the spin cut-off factor and $\omega(N,Z,U)$ is the state density. The state density can be deduced by inverse Laplace transform of the grand partition function:
\begin{equation} \label{11}
\omega(N,Z,U)=\frac{exp(S)}{{(2\pi)}^{3/2}|D|^{1/2}}.
\end{equation}
In the above equation, D is determinant of the second derivatives of the grand partition function taken at the saddle point.
For more information on the calculations of the formulas given in this section see our previous publications \cite{{razavi2012},{razavi2013},{razavi2014},{razavi ijmpe}}.

\section{Result and discussion}\label{result}
In the present work the single particle energies and their spins where first calculated by the using of Nilsson potential \cite{nilson potential}.The oscillator quantum number $\hbar\omega_0$ has been assigned the value of $41 A^{-\frac{1}{3}}$ MeV. The quantities $\mu$ and $\chi$, which enter in the Nilsson potential were taken from \cite{nilson parameters}.Obtained values have been used in BCS calculations for the number of neutrons and protons of $^{105,106,107}$Pd. Then the excitation energy, entropy and nuclear level density have been calculated by the use of Eqs.4,5 and 10, respectively.The neutron and proton pairing gap parameter at zero temperature, which were used to calculate pairing strength $G$ by solving Eq.2, where obtained by three point method \cite{3piont}.
Calculated nuclear level density for $^{106}$Pd in BCS model as a function of excitation energy is shown in Fig.1 beside corresponding experimental data. Examination of the figure shows that our results are in good agreement with the experimental data.
Using extracted values for entropy of $^{105,106,107}$Pd in BCS model, entropy excess have been calculated. Extracted single particle entropy and single hole entropy as a function of temperature are shown in Fig.2. Critical temperature is the temperature that gap parameter, that is a measure of nuclear pairing, becomes zero and the so-called phase transition occurs. According to Fig.2 before phase transition at $T=0.3 MeV$ both single particle and single hole cause equal entropy excess but after the critical temperature the entropy excess related to single particle is greater than single hole entropy. Also in the extracted entropies from experimental level densities this deference is observed (see Fig.3).
\begin{figure}[H]
\begin{center}
\includegraphics[width=8.6cm] {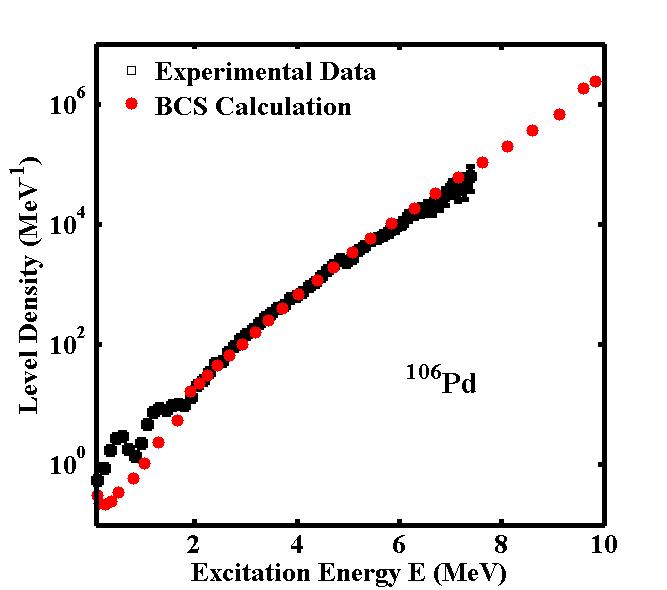}
\end{center}
\caption{The experimental \cite{exp.level density} and the calculated level density as a function of excitation energy in $^{106}$Pd nucleus.}
\end{figure}
\begin{figure}[H]
\begin{center}
\includegraphics[width=8.6 cm] {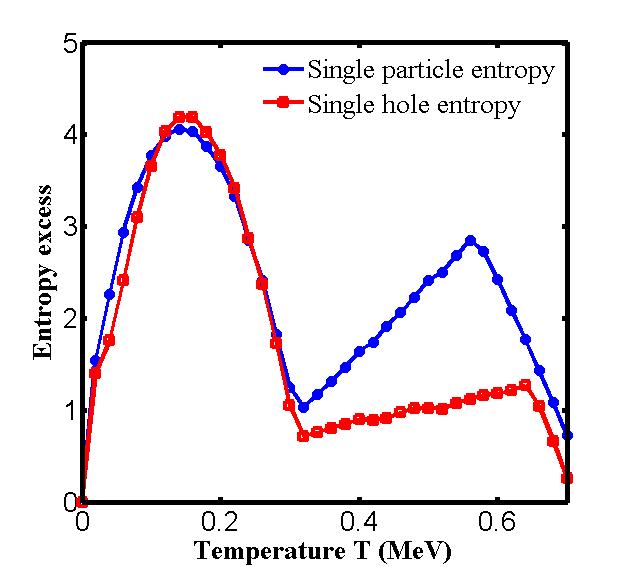}
\end{center}
\caption{Extracted entropy excess of $^{105}$Pd and $^{107}$Pd compared to $^{106}$Pd.}
\end{figure}

Using Eq.10 the role of neutron and proton system have been investigated for the entropy excess related to single particle and single hole. The results are shown in Fig.4 as a function of temperature. According to Fig.4 in both nuclei proton system have minor role in entropy excess before phase transition. In our previous work we showed that the proton system have minor role in entropy excess in $^{121}$Sn compared to $^{122}$Sn \cite{razavi2012}. According to this fact that the number of protons in Sn is a magic number $(Z=50)$ so protons system in $^{121}$Sn is expected to have stronger pairing correlations compared to Pd isotopes with $Z=46$. As we can see in Fig.4 increase in the role of proton system in entropy excess of $^{105}$Pd and $^{107}$Pd take place at lower temperature $(T=0.3 MeV)$ compared to $^{121}$Sn that is at $T=0.5 MeV$.

\begin{figure}[H]
\begin{center}
\includegraphics[width=8.6 cm] {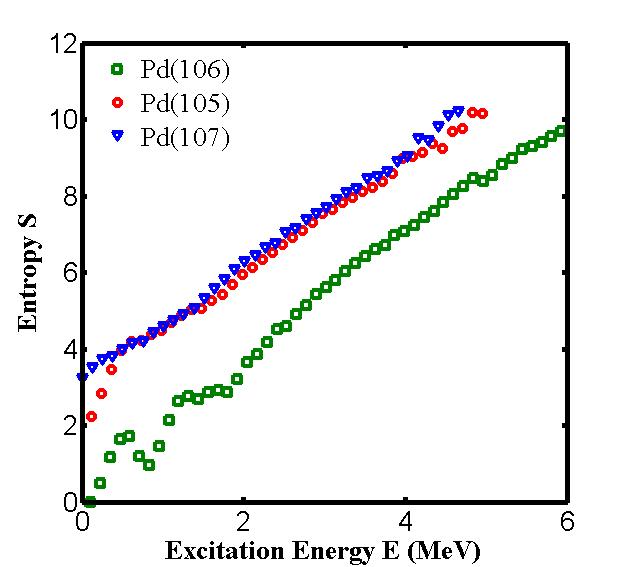}
\end{center}
\caption{Extracted entropies from experimental data on nuclear level densities of $^{105,106,107}$Pd .}
\end{figure}
\begin{figure}[H]
\begin{center}
\includegraphics[width=8.6 cm] {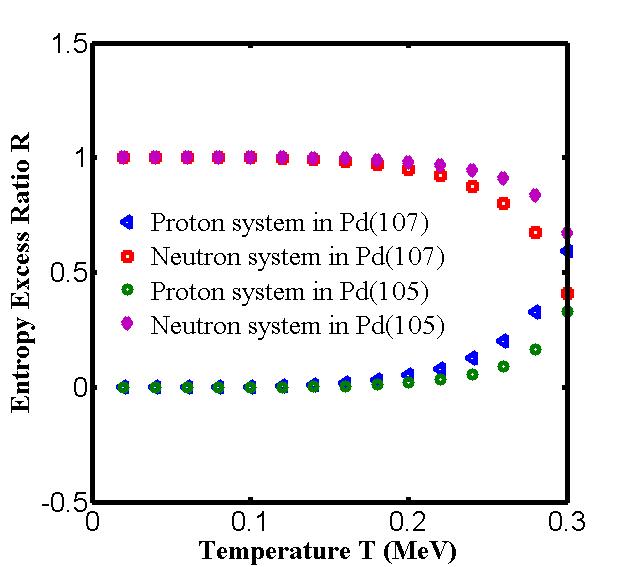}
\end{center}
\caption{Extracted entropy excess ratios for neutron and proton system of $^{105}$Pd and $^{107}$Pd compared to $^{106}$Pd.}
\end{figure}

\section{Conclusion}
In this work nuclear level densities and thermal quantities have been extracted for $^{105,106,107}$Pd within the BCS model. Then the entropy excess of $^{105}$Pd and $^{107}$Pd compared to $^{106}$Pd that interpreted as single hole and single particle entropy respectively have been extracted and their related curves as a function of temperature have been compared. The comparison shows that after phase transition the single particle entropy is greater than single hole entropy. This observation is in line with the observed behavior of extracted entropies from experimental level densities.
Using entropy excess ratio, the role of neutron and proton system in single hole and single particle entropy have been investigated. Our results show that in both cases proton system have minor role in entropy excess. Also, as the present results expected, proton system role in entropy of Pd isotopes increases at lower temperature compared to $^{121}$Sn with magic number $(Z=50)$.

\address{Department of Physics, Faculty of Science, University of Zanjan\\
 Zanjan, Iran\\
\email{a\_rahmatinejad@znu.ac.ir}}

\address{Department of Physics, Faculty of Science, Imam Khomeini International University\\
 Qazvin, Iran\\
\email{Tayeb@znu.ac.ir}}

\address{Department of Physics, Faculty of Science, Imam Hossein Comprehensive University\\
 Tehran, Iran\\
\email{rrazavin@ihu.ac.ir}}

\end{document}